\documentclass[10pt,aps,prd,nofootinbib,reprint,superscriptaddress]{revtex4-1}

\usepackage[utf8]{inputenc}
\usepackage{amsmath}
\usepackage{graphicx}
\usepackage{multirow}

\usepackage[colorlinks,pdfusetitle]{hyperref}
\hypersetup{colorlinks=true,allcolors=[rgb]{1,0.56,0}}

\DeclareMathOperator{\diag}{diag}

\newcommand{\orcid}[1]{\href{https://orcid.org/#1}{#1}}
\newcommand{\e}[1]{\times10^{#1}}
\newcommand{\eps}{\epsilon}

\begin{document}

\title{CP-Violating Neutrino Non-Standard Interactions in Long-Baseline-Accelerator Data}

\author{Peter B.~Denton}
\email{pdenton@bnl.gov}
\thanks{\orcid{0000-0002-5209-872X}}
\affiliation{High Energy Theory Group, Physics Department, Brookhaven National Laboratory, Upton, NY 11973, USA}

\author{Julia Gehrlein}
\email{jgehrlein@bnl.gov}
\thanks{\orcid{0000-0002-1235-0505}}
\affiliation{High Energy Theory Group, Physics Department, Brookhaven National Laboratory, Upton, NY 11973, USA}

\author{Rebekah Pestes}
\email{rebhawk8@vt.edu}
\thanks{\orcid{0000-0002-9634-1664}}
\affiliation{High Energy Theory Group, Physics Department, Brookhaven National Laboratory, Upton, NY 11973, USA}
\affiliation{Center for Neutrino Physics, Department of Physics, Virginia Tech, Blacksburg, VA 24061, USA}

\begin{abstract}
Neutrino oscillations in matter provide a unique probe of new physics.
Leveraging the advent of neutrino appearance data from NOvA and T2K in recent years, we investigate the presence of CP-violating neutrino non-standard interactions in the oscillation data.
We first show how to very simply approximate the expected NSI parameters to resolve differences between two long-baseline appearance experiments analytically.
Then, by combining recent NOvA and T2K data, we find a tantalizing hint of CP-violating NSI preferring a new complex phase that is close to maximal: $\phi_{e\mu}$ or $\phi_{e\tau}\approx3\pi/2$ with $|\eps_{e\mu}|$ or $|\eps_{e\tau}|\sim0.2$.
We then compare the results from long-baseline data to constraints from IceCube and COHERENT.
\end{abstract}

\maketitle

\section{Introduction}
Neutrino oscillations have provided the only particle physics evidence for new physics beyond the standard model (BSM) to date \cite{Fukuda:1998mi,Ahmad:2002jz}, making it an excellent place to probe new physics scenarios.
The phenomenology of neutrino oscillations is fairly unique, as it provides an opportunity to observe the accumulation of a relative phase over macroscopic distances, making neutrino oscillations one of the purest probes of quantum mechanics available.
During propagation, the environment may also modify the phases due to an interaction. Such an interaction exists in the standard model (SM) and is called the Wolfenstein matter effect \cite{Wolfenstein:1977ue}, wherein a neutrino in the electron state of the flavor basis experiences a potential with the background electrons via a charged-current (CC) interaction.

In the same paper that pointed out the SM matter effect, Wolfenstein also suggested the possibility of a new interaction that provides a matter effect, so-called neutrino non-standard interactions (NSI) \cite{Wolfenstein:1977ue,Farzan:2017xzy,Dev:2019anc}.
Since then, there has been an explosion of interest to probe these new interactions.
Numerous UV-complete models have been developed \cite{Forero:2016ghr,Denton:2018dqq,Dey:2018yht,Babu:2017olk,Farzan:2016wym,Farzan:2015hkd,Farzan:2015doa,Babu:2019mfe}\footnote{These models allow for sizable NSI via various mechanisms such as constraining the direct coupling of the NSI mediator to the heavier generations or to sterile neutrinos that mix with the active ones.} and the phenomenology has been generalized beyond vector currents \cite{Ge:2018uhz,AristizabalSierra:2018eqm,Bischer:2019ttk,Babu:2019iml}.
In addition, several NSI parameters introduce various interesting degeneracies in oscillation or scattering experiments \cite{Miranda:2004nb,GonzalezGarcia:2001mp,Kikuchi:2008vq,Coloma:2011rq,Friedland:2012tq,Rahman:2015vqa,Masud:2015xva,Coloma:2015kiu,Palazzo:2015gja,deGouvea:2015ndi,Masud:2016bvp,Liao:2016hsa,deGouvea:2016pom,Liao:2016orc,Ge:2016dlx,Agarwalla:2016fkh,Blennow:2016etl,Fukasawa:2016lew,Deepthi:2016erc,Forero:2016cmb,Farzan:2017xzy,Deepthi:2017gxg,Coloma:2017egw,Coloma:2017ncl,Hyde:2018tqt,Coloma:2019mbs,Esteban:2019lfo,Kopp:2010qt}, which demonstrates the importance of complementary measurements of the NSI parameters.

One of the most complete ways to probe neutrino oscillations is through long-baseline accelerator experiments with electron (anti)neutrino appearance.
While these measurements are extremely challenging experimentally, they provide a wealth of information, as they are sensitive to many oscillation parameters, including those that are the least constrained, like the CP-violating phase $\delta$ from the leptonic mass mixing matrix.
In addition, appearance measurements provide a crucial probe of certain NSI parameters.

The two state-of-the-art long-baseline neutrino experiments are NOvA and T2K \cite{Ayres:2007tu,Abe:2011ks}.
Both are off-axis; therefore, each detects a flux of neutrinos with a relatively narrow energy distribution.
The latest results from both experiments \cite{patrick_dunne_2020_3959558,alex_himmel_2020_3959581} show a slight tension at the $\sim2\sigma$ level, depending on how exactly it is quantified.
Both experiments prefer the normal mass ordering, but T2K prefers $\delta\sim3\pi/2$ while NOvA does not have much preference and is generally around $\delta\sim\pi$.
While this is not yet significant, it provides an interesting test case for new physics should it persist, as both experiments plan to accumulate additional data.

The parameters for this tension are particularly interesting.
Since a new physics explanation probably has to depend on the matter effect and since T2K with less matter effect sees some evidence of CPV, this means that CPV is present not only in the mass matrix but also in the new physics sector.
Thus we are presenting evidence of two cases of relatively ``large'' CPV in the neutrino sector.
Given the complex picture of CP with some parts of physics violating it maximally and others seeming to conserve it, these hints for extra CPV play a crucial role in our larger understanding of CP symmetry in physics.

In this paper, we review NSI and show how to approximate the NSI parameters that describe the NOvA and T2K data.
We then describe our treatment of the NOvA and T2K data.
Then, we show that the NOvA and T2K data can be resolved by the inclusion of NSI with complex CPV phases with a preference for CPV values over CP-conserving values.
Finally, we discuss our results in a broader picture of other neutrino measurements and present some possible plans to improve these results, and we conclude.\footnote{We also provide supplemental material for a derivation of some of our analytic results, our results in the standard oscillation picture, and some additional NSI results which includes \cite{Capozzi:2019iqn,1809173,Jarlskog:1985ht, Denton:2020igp, Gando:2013nba,Adey:2018zwh, Ayres:2004js,Itow:2001ee,Abe:2015zbg,Abe:2017aap,yasuhiro_nakajima_2020_3959640,Denton:2020exu,Ge:2019ldu,Mitsuka:2011ty}.}
All the relevant data files are available at \href{https://peterdenton.github.io/Data/NOvA+T2K_NSI/index.html}{peterdenton.github.io/Data/NOvA+T2K\_NSI/index.html}.

\section{NSI Overview}
\label{sec:NSI}
NSI in oscillations provides an additional contribution to the matter potential of the neutrino oscillation Hamiltonian in the weak basis
\begin{equation}
H=\frac1{2E}\left[U^\dagger
M^2
U+a
\begin{pmatrix}
1+\eps_{ee}&\eps_{e\mu}&\eps_{e\tau}\\
\eps_{e\mu}^*&\eps_{\mu\mu}&\eps_{\mu\tau}\\
\eps_{e\tau}^*&\eps_{\mu\tau}^*&\eps_{\tau\tau}
\end{pmatrix}
\right]\,,
\label{eq:HNSI}
\end{equation}
where $E$ is the neutrino energy, $U\equiv R_{23}(\theta_{23})U_{13}(\theta_{13},\delta)R_{12}(\theta_{12})$ is the PMNS mixing matrix \cite{Pontecorvo:1957cp,Maki:1962mu} that is parameterized in the usual way \cite{Tanabashi:2018oca}, $M^2\equiv\diag(0,\Delta m^2_{21},\Delta m^2_{31})$ is the diagonal mass-squared matrix, $a\equiv2\sqrt2G_FN_eE$ parameterizes the matter effect, and $N_e$ is the electron density.
The $\eps_{\alpha\beta}$ terms parameterize the size of the new interaction relative to the weak interaction and typically arise from effective Lagrangians of the form
\begin{equation}
\mathcal L_{\rm NSI}=-2\sqrt2G_F\sum_{\alpha,\beta,f}\eps_{\alpha\beta}^f(\bar\nu_\alpha\gamma^\mu\nu_\beta)(\bar f\gamma_\mu f)\,.
\label{eq:LNSI}
\end{equation}
For simplicity, we only consider NSI with vector mediators.
The Lagrangian level NSI parameters in eq.~\ref{eq:LNSI} are related to the Hamiltonian level terms in eq.~\ref{eq:HNSI} by $\eps_{\alpha\beta}=\sum_f\frac{N_f}{N_e}\eps_{\alpha\beta}^f$, where $N_f$ is the number density of fermion $f$.
In the context of oscillations, it isn't possible to identify which matter particles (electrons, up quarks, or down quarks) the new physics is coupled to without comparing neutrino trajectories through materials with different neutron fractions, such as the Earth and the sun.
Within the context of long-baseline trajectories through the crust, the neutron fraction is close to one.
While the NSI parameters are often taken to be real for simplicity, we consider complex NSI, where $\eps_{\alpha\beta}=|\eps_{\alpha\beta}|e^{i\phi_{\alpha\beta}}$ for $\alpha\neq\beta$, which violate CP \cite{GonzalezGarcia:2001mp,Gago:2009ij,Girardi:2014kca}, see ref.~\cite{GonzalezGarcia:2011my} for more on complex NSI.
Diagonal non-universal NSI \cite{Guzzo:1991hi} does not lead to CPV assuming CPT invariance and will not be considered here.

\begin{figure}
\centering
\includegraphics[width=\columnwidth]{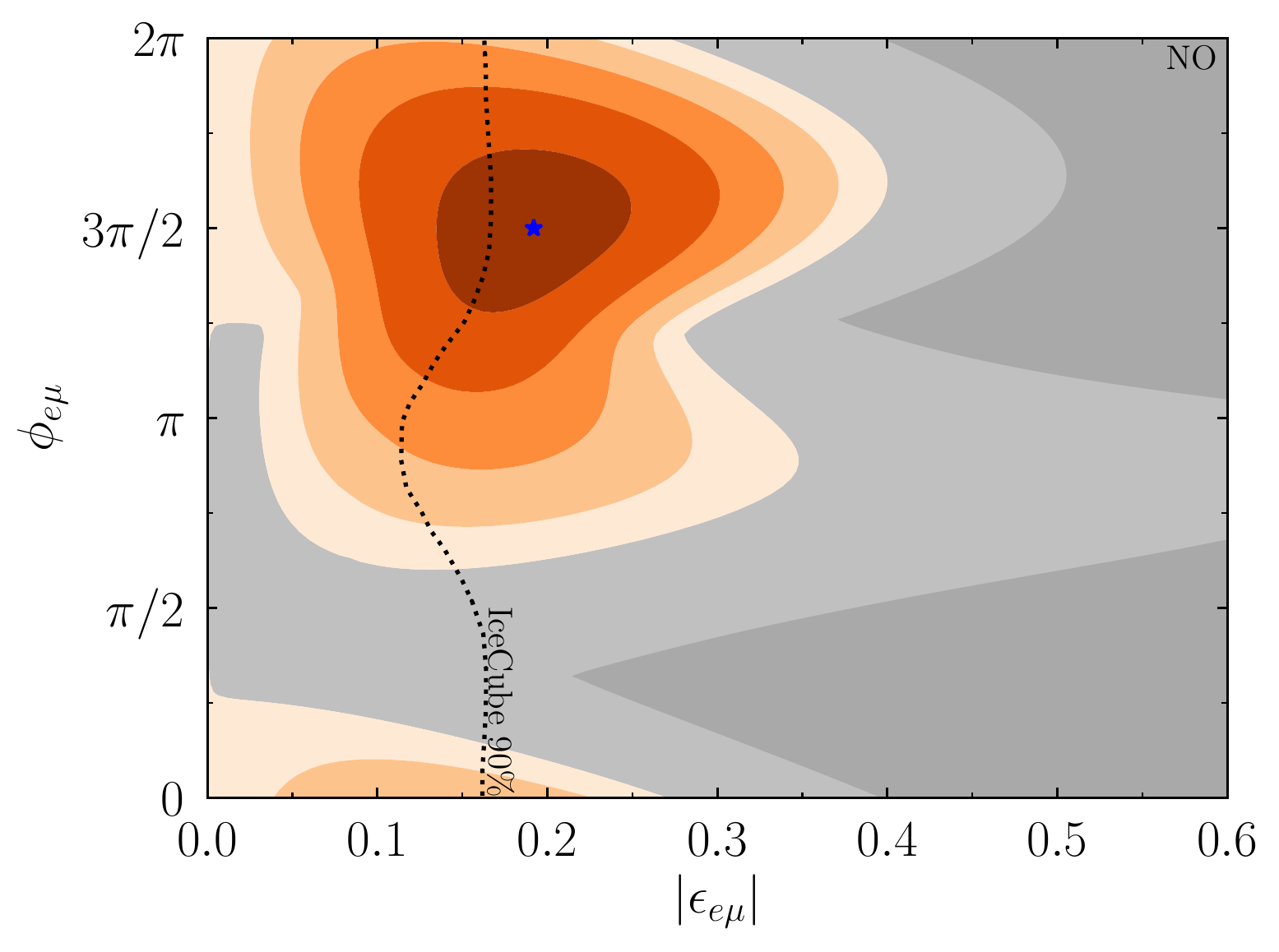}
\caption{The preferred parameter regions for $\epsilon_{e\mu}$ using the newest appearance and disappearance data from NOvA and T2K and assuming the NO.
The gray region is disfavored compared to the SM, and the dark gray region is ruled out by NOvA and T2K data at $\Delta\chi^2\le-4.61$.
The blue star shows the best fit point.
Each of the orange contours are drawn at integer values of $\Delta\chi^2$.
See table \ref{tab:bf} for the best parameters.
IceCube disfavors the region to the right of the black dotted curve at 90\% \cite{ICslides}.}
\label{fig:res_eps}
\end{figure}

One can analytically estimate the magnitude of the NSI parameter that would resolve different measurements of $\delta$ in experiments experiencing distinct matter potentials.
We find that if two experiments at two different matter potentials measure two disparate values of $\delta$ due to $\eps_{e\beta}$ NSI for $\beta\in\{\mu,\tau\}$, the magnitude of the NSI in the normal ordering (NO) is approximately given by
\begin{align}
|\eps_{e\beta}|&\approx\frac{s_{12}c_{12}c_{23}\pi\Delta m^2_{21}}{2s_{23}w_\beta}\left|\frac{\sin\delta_{\rm T2K}-\sin\delta_{\rm NOvA}}{a_{\rm NOvA}-a_{\rm T2K}}\right|
\label{eq:eps approx}\\
&\approx
\begin{cases}
0.22&\text{for }\beta=\mu\\
0.24\qquad&\text{for }\beta=\tau
\end{cases}\,,\nonumber
\end{align}
where $w_\beta=s_{23}$ or $c_{23}$ for $\beta=\mu$ or $\tau$ respectively\footnote{We use the standard $c_{ij}=\cos\theta_{ij}$, $s_{ij}=\sin\theta_{ij}$ shorthand.}, see the supplemental material for a derivation.
The preferred value of $\eps_{e\tau}$ is larger than that for $\eps_{e\mu}$ since T2K prefers the upper octant and T2K is less affected by NSI than NOvA.
The difference between $\eps_{e\mu}$ and $\eps_{e\tau}$ makes sense since long-baseline oscillations are dominated by $\nu_3$, which contains more $\nu_\mu$ in the upper octant, and thus not as much NSI affecting $\nu_\mu$ is required to produce a given effect.
We also note that the approximations presented here are quite consistent with our numerical results discussed below and shown in fig.~\ref{fig:res_eps} and table \ref{tab:bf}.
In addition, in some regions of parameter space, it may be possible to connect $\delta$ and the NSI phases via a technique known as phase reduction \cite{Kikuchi:2008vq, Ribeiro:2007ud}.

\section{Analysis Details}
\label{sec:analysis}
The appearance channels at NOvA and T2K can be approximated by counting experiments, while for the disappearance channels, the energy distribution of the events is important.
This approximation ignores several potentially problematic issues: the energy distributions aren't exactly delta distributions, there are correlated systematics between the different channels, and the cross section systematics may well be related even between the different experiments.
Nonetheless, we find an acceptable reproduction of the results with the simple treatment described below.

NOvA measures neutrinos with $E\sim1.9$~GeV after traveling 810~km through the Earth with density $\rho=2.84$~g/cc, while
T2K measures neutrinos with $E=0.6$~GeV after traveling 295~km through the Earth with average density  $\rho=2.6$~g/cc.
For the appearance channels, we find that the number of events can be expressed as a constant normalization term and a constant factor which multiplies the oscillation probability in matter (see also \cite{Kelly:2020fkv} for a similar approach).
These constant factors can be derived from the provided bi-event plots in \cite{alex_himmel_2020_3959581,novaichep, patrick_dunne_2020_3959558}. As wrong sign leptons contribute to the flux, especially in antineutrino mode, we parameterize the predicted numbers of events as 
\begin{align}
n(\nu_e)=x P(\nu_\mu\to \nu_e) +y P(\bar{\nu}_\mu\to \bar{\nu}_e) + z\,,
\end{align}
and similarly for the antineutrino channel, where $x,~y,~z$ are real numbers which roughly translate to the weighted neutrino (antineutrino) flux times cross-section  for  this  particular  energy, and the background rate in this channel\footnote{Unlike other recent analyses of NOvA and T2K data, we include the wrong sign lepton contribution as it considerably improves our description of the experiment.}.
For NOvA, a good fit is obtained for the neutrino channel without including the wrong sign leptons,
\begin{align}
n(\nu_e)^{\text{NOvA}}={}&31.15+1149.7\times P(\nu_\mu\to \nu_e)~,\\
n(\bar{\nu}_e)^{\text{NOvA}}={}&13.97+
472.60\times P(\bar{\nu}_\mu\to \bar{\nu}_e)\nonumber\\&+22.96\times P(\nu_\mu\to \nu_e)\,,
\end{align}
while for T2K, we find
\begin{align}
n(\nu_e)^{\text{T2K}}={}&
19.71+ 1284.16 \times P(\nu_\mu\to \nu_e)\nonumber\\&+ 36.90 \times P(\bar{\nu}_\mu\to \bar{\nu}_e)~,\\
n(\bar{\nu}_e)^{\text{T2K}}={}&5.84 +231.32 \times P(\bar{\nu}_\mu\to \bar{\nu}_e)\nonumber\\&+
49.51 \times P(\nu_\mu\to \nu_e)~.
\end{align}

At leading order, the oscillation probability for neutrinos and antineutrinos is the same for the disappearance channel. However, this changes in the presence of NSI. In the following, we will assume that the results in the disappearance channel are dominated by the neutrino sample, which provides higher statistics than the antineutrino sample.
We adapt the results from \cite{Kelly:2020fkv} for the disappearance channel at NOvA, where they found as best fit
$|\Delta m_{32}^2|=(2.41\pm0.07)\times10^{-3}~\text{eV}^2$ and $4|U_{\mu3}|^2(1- |U_{\mu3}|^2) = 0.99\pm 0.02$.
For T2K, we obtain the test statistic for $\theta_{23}$ and $\Delta m^2_{32}$ from the 1D distributions of the test statistics provided by the experiment \cite{patrick_dunne_2020_3959558}\footnote{While these distributions do include information from the appearance mode, we assume that they are dominated by the high statistics measurements made in disappearance mode.}. 

For the appearance channel, incorporating the effect of NSIs as described in eq.~\ref{eq:HNSI} is straightforward.
For the disappearance channels, we calculate the effective vacuum mixing parameters by solving
\begin{equation}
U^\dagger M^2U+A+N=\widetilde U^\dagger\widetilde M^2\widetilde U+A\,,
\label{eq:nsi to effective parameters}
\end{equation}
where $A\equiv\diag(a,0,0)$ and the $N$ matrix contains the $\eps$'s and is proportional to the matter potential $a$.
Then, by diagonalizing $U^\dagger M^2U+N$, one finds the vacuum parameters that a long-baseline accelerator experiment would extract in the presence of NSI at a given energy.
Various approximate techniques for the diagonalization of matrices in the context of neutrino oscillations in matter have been explored in \cite{Yokomakura:2000sv,Kimura:2002wd,Agarwalla:2013tza,Minakata:2015gra,Denton:2016wmg,Denton:2018hal,Denton:2018fex,Denton:2019qzn}.
The approach presented in eq.~\ref{eq:nsi to effective parameters} is exact in the case of constant matter density; it does not apply to solar or atmospheric neutrinos, and additional care is necessary there.
Finally, one can compare the effective vacuum mixing parameters extracted from $\widetilde M^2$ and $\widetilde U$ to the measured oscillation parameters.

To analyze the data, we construct a test statistic using a log likelihood ratio with Poisson statistics for the appearance data and simple $\chi^2$ pulls for the disappearance constraints.
We show the results in the standard oscillation picture in the supplemental material, which show the preference for the IO when the two experiments are combined without NSI.

In the next section, we find that, in the presence of NSI however, the long-baseline data is better described by the NO than the IO, so we assume the the true mass ordering is normal, unless otherwise specified.
This is crucial as the mass ordering affects many other experiments including end point, neutrinoless-double-beta decay, and cosmological measurements.
The MO can be confirmed independently of the presence of NSI via JUNO \cite{Djurcic:2015vqa}.

\begin{table}
\centering
\caption{Best fit values and $\Delta\chi^2=\chi^2_{\rm SM}-\chi^2_{\rm NSI}$ for a fixed MO considering one complex NSI parameter at a time. (For the SM, $\chi^2_{\rm NO}-\chi^2_{\rm IO}=2.3\,$.)}
\begin{tabular}{c|c||c|c|c|c}
MO&NSI&$|\eps_{\alpha\beta}|$&$\phi_{\alpha\beta}/\pi$&$\delta/\pi$&$\Delta\chi^2$\\\hline
\multirow{3}{*}{NO}&$\eps_{e\mu}$&0.19&1.50&1.46&4.44\\
&$\eps_{e\tau}$&0.28&1.60&1.46&3.65\\
&$\eps_{\mu\tau}$&0.35&0.60&1.83&0.90\\\hline
\multirow{3}{*}{IO}&$\eps_{e\mu}$&0.04&1.50&1.52&0.23\\
&$\eps_{e\tau}$&0.15&1.46&1.59&0.69\\
&$\eps_{\mu\tau}$&0.17&0.14&1.51&1.03
\end{tabular}
\label{tab:bf}
\end{table}

\section{NSI Results}
\label{sec:NSI results}
We analyze one complex NSI parameter at a time, using the appearance and disappearance data from NOvA and T2K and assuming the NO.
In fig.~\ref{fig:res_eps}, we present the allowed parameter regions in the $|\eps_{\alpha\beta}|$-$\phi_{\alpha\beta}$ plane for $\eps_{e\mu}$.
The results for $\eps_{e\tau}$ and $\eps_{\mu\tau}$ can be found in the supplemental material.
For simplicity, we fix $\theta_{13}$, $\theta_{12}$, and $\Delta m^2_{21}$ to the best fit values from Daya Bay and KamLAND as described above, as these experiments are not affected by NSI\footnote{The slight discrepancy between the determination of the solar mass splitting by KamLAND and SuperK could be resolved by NSI in the Sun \cite{Liao:2017awz}.}, and marginalize over $\Delta m^2_{31}$, $\delta$, and $\theta_{23}$, including the pull on $\Delta m^2_{31}$ from Daya Bay.
We have verified that including the pulls associated with $\theta_{13}$, $\theta_{12}$, and $\Delta m^2_{21}$ do not significantly affect our results.
The best fit values for the parameters for each case of $\eps_{e\mu}$, $\eps_{e\tau}$, and $\eps_{\mu\tau}$ in both MOs are given in table \ref{tab:bf}.
Note that while the combination of both experiments raises the $\chi^2$ by about 5.5 as mentioned in the supplemental material, that can be nearly completely alleviated with the addition of $\eps_{e\mu}$ which provides an improvement in the test statistic of 4.44 (compare this to switching to the IO which only improves the test statistic by 2.3 and is in tension with SK data).
In the presence of NSI, we still find that the upper octant is preferred with $\sin^2\theta_{23}=0.56$ for all three NSI parameters and both MOs.

Consistent with our analytic estimates, we find moderate evidence for CP-violating NSI.
The best solution is with the $\eps_{e\mu}$ parameter with maximal CP-violating phases for both the standard CP phase and the new NSI CP phase.

Other oscillation probes of NSI come from atmospheric and solar experiments.
As atmospheric constraints are expected to be stronger than those from solar, we focus on those.
The constraints on complex NSI parameters from IceCube \cite{ICslides} slightly disfavor the preferred region for $\eps_{e\mu}$, although it is possible to get an improved fit to the NOvA and T2K data while not being in too strong of tension with the IceCube data. In fact, the best fit point to the IceCube data for $\eps_{e\mu}$ is at $|\eps_{e\mu}|=0.07$ and $\phi_{e\mu}/\pi=1.91$, close to the relevant numbers for NOvA and T2K.
It is also interesting to note that IceCube slightly disfavors $|\eps_{e\mu}|=0$ at just over $1\sigma$.

We show the constraints from IceCube on complex NSI from \cite{ICslides} on fig.~\ref{fig:res_eps} and in the supplemental material, which only slightly disfavors this NSI explanation of long-baseline data with $\eps_{e\mu}$.
The IceCube constraints are comparable to other constraints in the literature on real NSI from oscillation experiments \cite{Coloma:2017egw,Esteban:2018ppq,Farzan:2017xzy}.
Constraints on complex off-diagonal NSI from solar measurements are expected to be weak \cite{Miranda:2004nb}.

COHERENT's measurement of the coherent elastic neutrino nucleus scattering (CEvNS) process \cite{Freedman:1973yd} provides constraints \cite{Akimov:2017ade,Akimov:2020pdx,Coloma:2017egw,Denton:2020hop,Coloma:2019mbs,Liao:2017uzy,Miranda:2020tif,AristizabalSierra:2018eqm,Giunti:2019xpr,Papoulias:2019xaw} on the NSI parameter space that is also an explanation of the NOvA and T2K data.
Further constraints come from elastic neutrino electron scattering \cite{Khan:2017oxw}.
While the parameters relevant for NOvA and T2K are not strongly ruled out by scattering experiments yet \cite{Denton:2020hop,Dutta:2020che}, they can be probed by COHERENT in coming years. 
It should be noted, however, that the NSI constraint derived from COHERENT only applies to NSI governed by mediators heavier than $\sim10$ MeV \cite{Coloma:2017egw,Denton:2018xmq}.
Constraints for lower mediator masses down to $\sim1$ MeV can by placed with upcoming low-threshold CEvNS experiments at nuclear reactors.
Meanwhile, early universe measurements constrain mediators lighter than $\sim5$ MeV \cite{Kamada:2015era,Huang:2017egl}.
Thus we anticipate that COHERENT or future reactor CEvNS experiments should be able to probe the NSI parameters that could explain the NOvA and T2K data in coming years.

\section{Conclusions}
\label{sec:conclusions}
Measuring and understanding CP violation is of the utmost importance in particle physics.
Somewhat confusingly, the weak interaction violates CP while the strong interaction seems to conserve CP.
Meanwhile, the quark mass mixing matrix has relatively small CP violation.
To better understand the important role that CPV plays in particle physics, we must measure it and understand it in the leptonic sector.

In this manuscript, we have analyzed a new physics explanation for the slight tension in the recent NOvA and T2K data.
We performed a fit to the data and showed that this tension can be resolved when introducing complex CP-violating NSI parameters. As an example, we analyzed non-zero $\eps_{e\mu},~\eps_{e\tau},~\eps_{\mu\tau}$ one at a time and found that the best fit points for the new complex phases of $\eps_{\alpha\beta}$ prefers not only maximal CPV in the new interaction around $3\pi/2$ for $\alpha=e$, but also large CPV in the leptonic mass matrix.
These NSI parameters are best constrained (not counting long-baseline experiments) by atmospheric oscillation measurements by Super-KamiokaNDE and IceCube.
These measurements rule out the favored parameter region for $\eps_{\mu\tau}$, whereas the atmospheric constraints only partially disfavor the preferred regions of $\eps_{e\mu}$ and $\eps_{e\tau}$.
We anticipate that improvements from Super-KamiokaNDE and IceCube can further test this hypothesis in the future\footnote{Future long-baseline experiments also have improvedsensitivity to the range of NSI parameters considered here \cite{Liao:2017awz,deGouvea:2015ndi}.}.
Furthermore, experiments that probe coherent elastic neutrino nucleus scattering will provide strong constraints on NSI parameters of a similar order of magnitude, though they currently only apply to mediators heavier than the $\sim10$ MeV scale.

The connections between combining experiments, the mass ordering, and NSI lead to this narrative:
\begin{enumerate}
\item Without new physics, NOvA and T2K each individually prefer the NO.
\item Their combination, without new physics, slightly prefers the IO over NO, despite \#1 above.
\item When allowing for CPV NSI, the preference is for the new physics in the NO over the standard oscillation picture.
\end{enumerate}
We also point out that JUNO's measurement of the MO, which has almost no dependence on the matter effect, will determine the MO independent of NSI.

We can see clearly from e.g.~eq.~\ref{eq:nsi to effective parameters} that in order to measure NSI with long-baseline neutrinos, one needs to either compare two different experiments or use a broad band beam such as that which DUNE will have \cite{Acciarri:2015uup}.
If this hint for CPV NSI persists, T2HK will find a similar value for $\delta$ as T2K has, while DUNE should be able to see some evidence for NSI directly.

To summarize, we have shown that the tension of the recent NOvA and T2K data can be resolved in a BSM scenario with the introduction of CP-violating NSI parameters, which can be further probed with near-future experiments.
It would be interesting to see if other new physics models could also explain the discrepancy, such as the presence of sterile neutrinos, decoherence, or neutrino decay.

\begin{acknowledgments}
We acknowledge support from the US Department of Energy under Grant Contract DE-SC0012704.
The work presented here that RP did was supported by the U.S.~Department of Energy, Office of Science, Office of Workforce Development for Teachers and Scientists, Office of Science Graduate Student Research (SCGSR) program.
The SCGSR program is administered by the Oak Ridge Institute for Science and Education (ORISE) for the DOE. 
ORISE is managed by ORAU under contract number DE-SC0014664. All opinions expressed in this paper are the authors' and do not necessarily reflect the policies and views of DOE, ORAU, or ORISE.
\end{acknowledgments}

\bibliography{main}

\appendix
\onecolumngrid

\section{Analytic Derivation}
\label{sec:analytic}
Since the inclusion of NSI allows one, in principle, to exactly map one set of vacuum parameters onto another (see eq.~9 in the main text), we can write down a system of equations of the form
\begin{align}
P(\eps=0,\delta_{\rm meas})&=P(\eps,\delta_{\rm true})\,,\\
\bar P(\eps=0,\delta_{\rm meas})&=\bar P(\eps,\delta_{\rm true})\,,
\end{align}
where we require both neutrino and antineutrino modes are equal for a given experiment. For simplicity, we assume that the effect of NSI is completely absorbed in the CP phase; in principle, the other parameters are also altered by NSI, specifically $\theta_{23}$ and $\Delta m^2_{31}$, but we assume that the impact of NSI on those parameters is small, as will be justified by comparing our analytic and numerical results.
Here, $\delta_{\rm meas}$ is the value of $\delta$ extracted by the experiment when assuming the standard oscillation picture.
That is, the LHS represents the probabilities as a function of the parameters extracted, assuming no new physics, while the RHS represents the probabilities in terms of the ``true'' parameters.

We can use approximate expressions for NSI in long-baseline experiments to determine the relationship among the measured values of $\delta$, the true value of $\delta$, and the magnitude and phase of the NSI.
From refs.~\cite{Kikuchi:2008vq,Capozzi:2019iqn} after some manipulation, we find, for neutrinos and antineutrinos respectively,
\begin{align}
-s_{12}c_{12}c_{23}\frac\pi2\Delta m^2_{21}\sin\delta+a_{\rm NOvA}|\eps_{e\beta}|\left[w_\beta s_{23}\cos(\delta+\phi_{e\beta})-v_\beta c_{23}\frac\pi2\sin(\delta+\phi_{e\beta})\right]&\approx-s_{12}c_{12}c_{23}\frac\pi2\Delta m^2_{21}\sin\delta_{\rm NOvA}\,,\\
s_{12}c_{12}c_{23}\frac\pi2\Delta m^2_{21}\sin\delta-a_{\rm NOvA}|\eps_{e\beta}|\left[w_\beta s_{23}\cos(\delta+\phi_{e\beta})+v_\beta c_{23}\frac\pi2\sin(\delta+\phi_{e\beta})\right]&\approx s_{12}c_{12}c_{23}\frac\pi2\Delta m^2_{21}\sin\delta_{\rm NOvA}\,,
\end{align}
where $w_\beta=s_{23}$ ($c_{23}$), $v_\beta=c_{23}$ ($-s_{23}$)
for $\beta=\mu$ ($\tau$), and we have assumed that the NO is correct and that both experiments measure the NO.
A similar expressions exists for T2K, as well.
The fact that the only true phase that appears in these approximations is $\delta+\phi_{e\beta}$ is connected to the concept of phase reduction \cite{Kikuchi:2008vq}.

From the requirement that the probabilities in the neutrino and antineutrino channel should both be satisfied with the same parameters, one immediately finds that $\sin(\delta+\phi_{e\beta})=0$.
This means $\delta+\phi_{e\beta}=0$ or $\pi$ and that either $\cos(\delta+\phi_{e\beta})=1$ or $\cos(\delta+\phi_{e\beta})=-1$, respectively.
Plugging this in and subtracting the NOvA and T2K equations, we find
\begin{equation}
|\eps_{e\beta}|\approx\frac{s_{12}c_{12}c_{23}\pi\Delta m^2_{21}(\sin\delta_{\rm T2K}-\sin\delta_{\rm NOvA})}{2s_{23}w_\beta(a_{\rm NOvA}-a_{\rm T2K})\cos(\delta+\phi_{e\beta})}\,.
\end{equation}
Given that $a_{\rm NOvA}>a_{\rm T2K}$ and that the data suggests that $\sin\delta_{\rm T2K}<\sin\delta_{\rm NOvA}$, we find that $\cos(\delta+\phi_{e\beta})=-1$, and thus, $\delta+\phi_{e\beta}=\pi$.
In any case, we can write down the general result using absolute values, as shown in eq.~3 in the main text.
We obtain,
\begin{align}
|\eps_{e\beta}|&\approx\frac{s_{12}c_{12}c_{23}\pi\Delta m^2_{21}}{2s_{23}w_\beta}\left|\frac{\sin\delta_{\rm T2K}-\sin\delta_{\rm NOvA}}{a_{\rm NOvA}-a_{\rm T2K}}\right|
\approx
\begin{cases}
0.22&\text{for }\beta=\mu\\
0.24\qquad&\text{for }\beta=\tau
\end{cases}\,,
\end{align}
where we plugged in the numbers for the last part which result in NSI values generally consistent with those from the exact numerical searches.

We can instead divide the NOvA and T2K equations to find
\begin{equation}
\sin\delta\approx\frac{\sin\delta_{\rm NOvA}a_{\rm T2K}-\sin\delta_{\rm T2K}a_{\rm NOvA}}{a_{\rm T2K}-a_{\rm NOvA}}\,.
\label{eq:sind approx}
\end{equation}
Plugging in the numbers, we find that the true value of $\delta$ one would expect is $\sin\delta=-1.7$ (the unphysicality of this is due to our $\theta_{23}$ approximation, but it does not mischaracterize the general features of these approximations).
This means that for an NSI explanation of NOvA and T2K, we would expect $\sin\delta=-1$, and T2K would infer $\sin\delta_{\rm T2K}$ slightly larger than $-1$.
In addition, the effect of eq.~\ref{eq:sind approx} in our situation of $\sin\delta_{\rm T2K}\sim-1$ is somewhat alleviated by changes in $\theta_{23}$ due to NSI which we have not accounted for.
Given that we have $\cos(\delta+\phi_{e\beta})=-1$ in our scenario, in the limit where $a_{\rm T2K}\to0$, we see from eq.~\ref{eq:sind approx} that $\sin\delta\approx\sin\delta_{\rm T2K}$ as expected and that $\delta_{\rm T2K}+\phi_{e\beta}=\pi$, and thus $\phi_{e\beta}=3\pi/2$, consistent with our numerical results.

All of these results are derived assuming the approximate expressions from ref.~\cite{Capozzi:2019iqn}, that the experiments are at the first oscillation maximum, and that the matter potentials are small relative to $\Delta m^2_{31}$ (for NOvA (T2K) we have $a/\Delta m^2_{31}\approx1/6$ ($1/20$)).

\section{Standard Oscillation Results}
\label{sec:SM results}
\begin{figure}
\centering
\includegraphics[width=0.49\columnwidth]{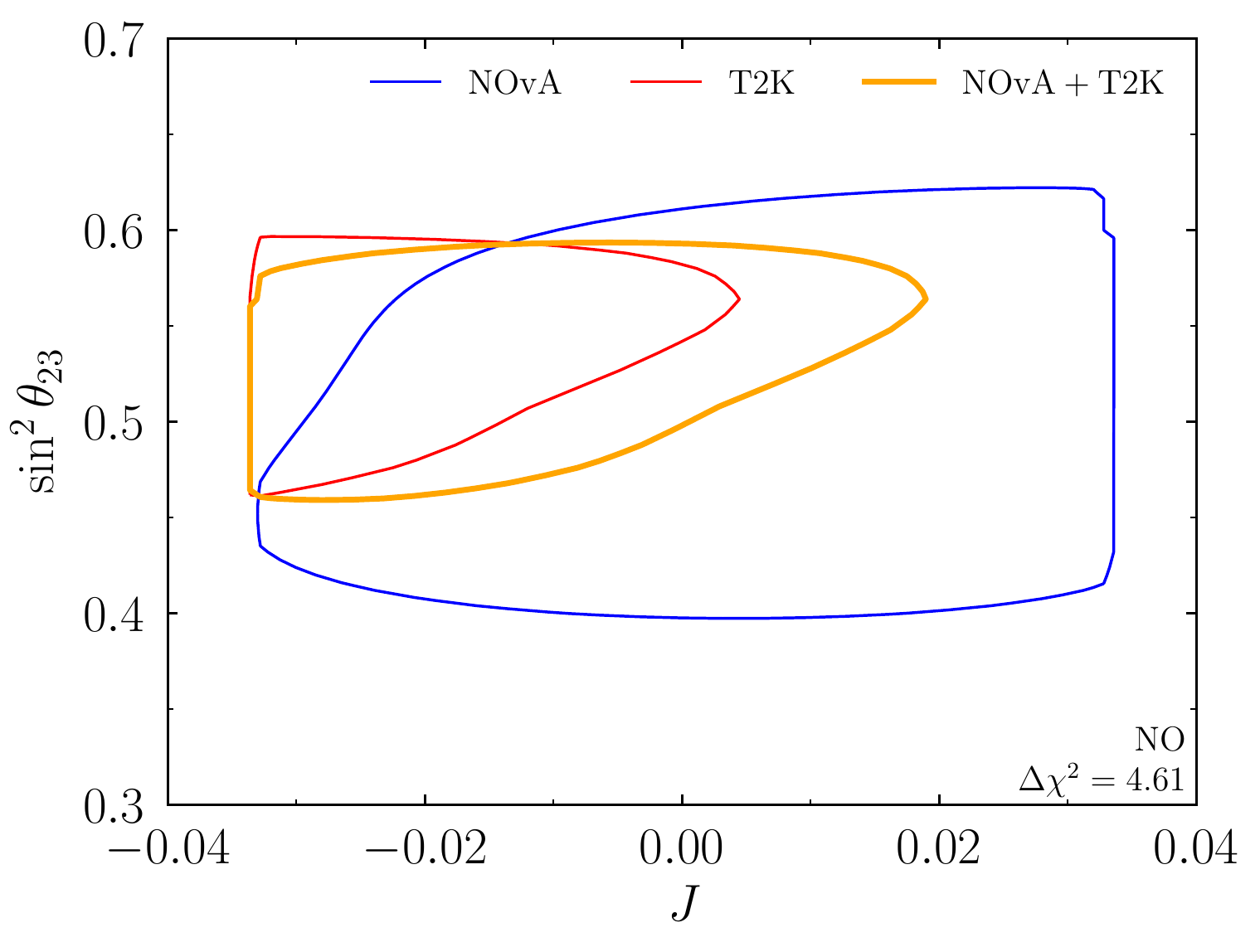}
\includegraphics[width=0.49\columnwidth]{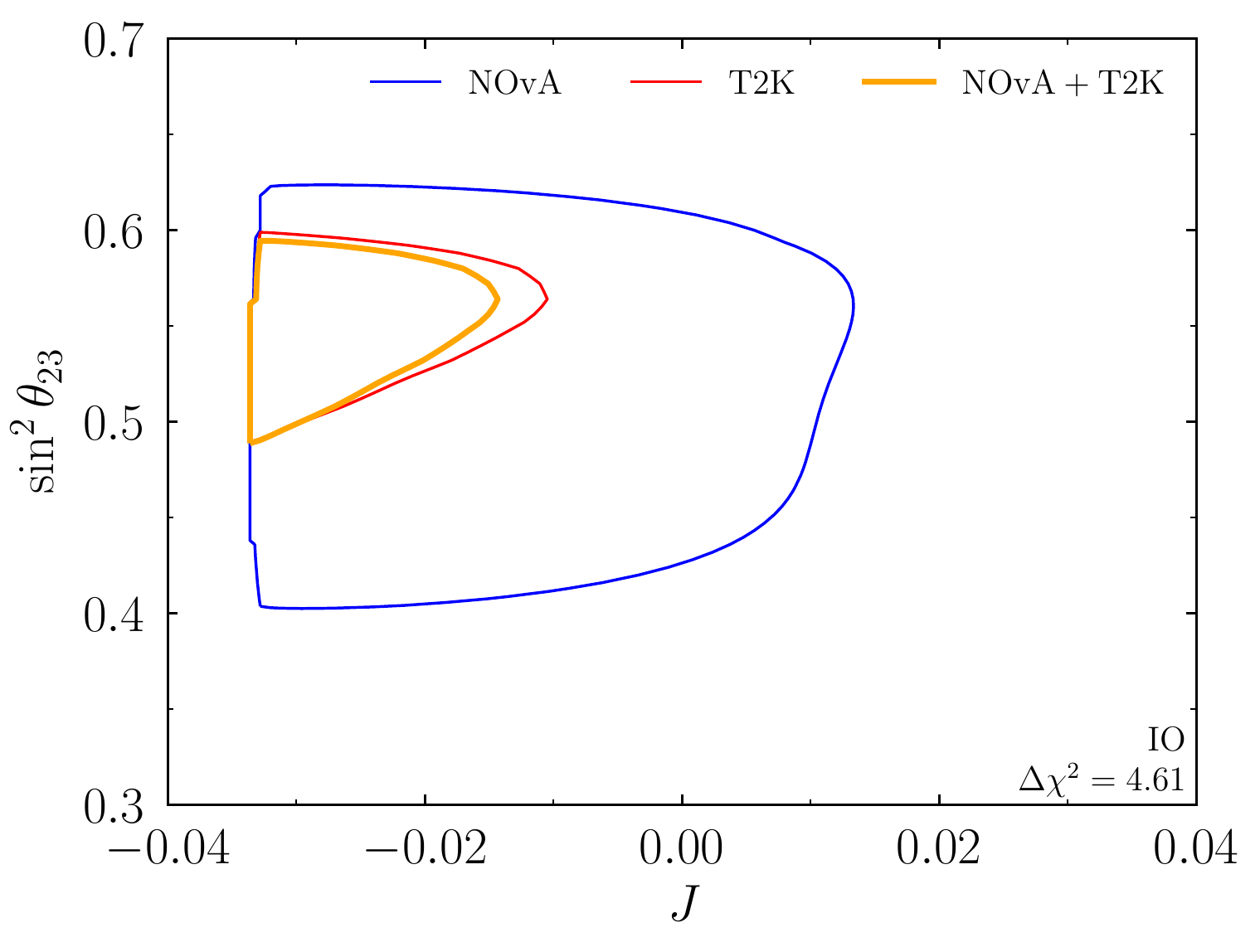}
\caption{The preferred regions in $\sin^2\theta_{23}$-$J$ space for the NO (left) and IO (right) for NOvA data, T2K data, or their combination at $\Delta\chi^2=4.61$ within the standard oscillation picture.
This includes a marginalization over $\Delta m^2_{21}$, $\Delta m^2_{31}$, $\theta_{13}$, $\theta_{12}$, and the sign of $\cos\delta$ with pulls from KamLAND and Daya Bay.}
\label{fig:SM}
\end{figure}

In addition to addressing new physics in the neutrino sector, we also show the preferred regions in the standard oscillation picture in fig.~\ref{fig:SM}.
Contours are drawn relative to the best fit point at $\Delta\chi^2=\chi^2-\chi^2_{\rm bf}=4.61$.
Note that combining the data sets within the normal mass ordering (NO) raises the minimum $\chi^2$ by $\sim5.5$ over either experiment individually; this tension can be somewhat alleviated by switching the mass ordering \cite{Kelly:2020fkv,1809173}.
We show the preferred regions of $\theta_{23}$ and the Jarlskog invariant where $J=s_{12}c_{12}s_{13}c_{13}^2s_{23}c_{23}\sin\delta$ is the Jarlskog \cite{Jarlskog:1985ht}, which is a parameterization-independent quantification of CPV in the leptonic mass matrix \cite{Denton:2020igp}.
Note that the maximum value of the Jarlskog is $1/6\sqrt3\approx0.096$; we are already quite far from maximal CPV in the leptonic sector due primarily to the fact that $\theta_{13}$ is fairly small.

For fig.~\ref{fig:SM} we include a minimization over the four other standard oscillation parameters and the sign of $\cos\delta$ for the Jarlskog panel.
We include priors from KamLAND \cite{Gando:2013nba} $\tan^2\theta_{12}=0.436^{+0.029}_{-0.025}$ and $\Delta m^2_{21}=(7.53\pm0.18)\e{-5}$ eV$^2$ as well as from Daya Bay \cite{Adey:2018zwh} $\sin^22\theta_{13}=0.0856\pm0.0029$ and $\Delta m^2_{32}=(2.471^{+0.068}_{-0.070})\e{-3}$ eV$^2$.
We find that the best fit parameters are at $J=-0.0120$, $\delta/\pi=1.21$, and $\sin^2\theta_{23}=0.556$ in the NO.
In the inverted mass ordering (IO) the best fit parameters are $J=-0.0328$, $\delta/\pi=1.54$, and $\sin^2\theta_{23}=0.560$.
These are compatible at the $<1\sigma$ level with the latest global fit to all oscillation experiments \cite{1809173}.

We see that in the NO, while T2K has some significance to disfavor $J=0$, the inclusion of NOvA data weakens this, making CPV in the standard oscillation picture an important goal for NOvA and T2K \cite{Ayres:2004js,Itow:2001ee} in coming years, as well as upcoming long-baseline accelerator neutrino experiments such as DUNE and T2HK \cite{Acciarri:2015uup,Abe:2015zbg}.
This weakening of the significance in the NO when the experiments are combined emphasizes the slight tension between the experiments.

Similarly to refs.~\cite{1809173,Kelly:2020fkv}, we also find that while NOvA and T2K both individually prefer the NO, the combination shows a slight preference for the IO at $\chi^2_{\rm NO}-\chi^2_{\rm IO}=2.3$.
When combined with Super-KamiokaNDE (SK) atmospheric data \cite{Abe:2017aap,yasuhiro_nakajima_2020_3959640}, the best fit mass ordering (MO) remains normal \cite{1809173,Kelly:2020fkv}\footnote{SK preferred the NO at $\chi^2_{\rm IO}-\chi^2_{\rm NO}>5$, but with their latest data release, the significance dropped to $\sim3.2$, although it is still enough to prefer the NO in total.}.
This MO question is of crucial significance beyond just measuring parameters in the SM.
It may provide guidance about the structure of neutrino mass \cite{Denton:2020exu} and is a key input for many experimental measurements of neutrinos, including cosmological measurements of neutrino properties, kinematic measurements of neutrinos, and neutrinoless-double-beta decay measurements should neutrinos have a Majorana mass term, see e.g.~\cite{Ge:2019ldu}.

We see that in the NO the allowed region for both experiments is larger than that for T2K which highlights the tension between the experiments in the NO.
On the other hand, in the IO the allowed region for both experiments is smaller than that for either experiment which shows that both experiments find slightly better agreement in the IO than in the NO.

\section{Results for \texorpdfstring{$\eps_{e\tau}$}{eps e tau} and \texorpdfstring{$\eps_{\mu\tau}$}{eps mu tau}}
\label{sec:epsmt}
\begin{figure}
\centering
\includegraphics[width=0.48\linewidth]{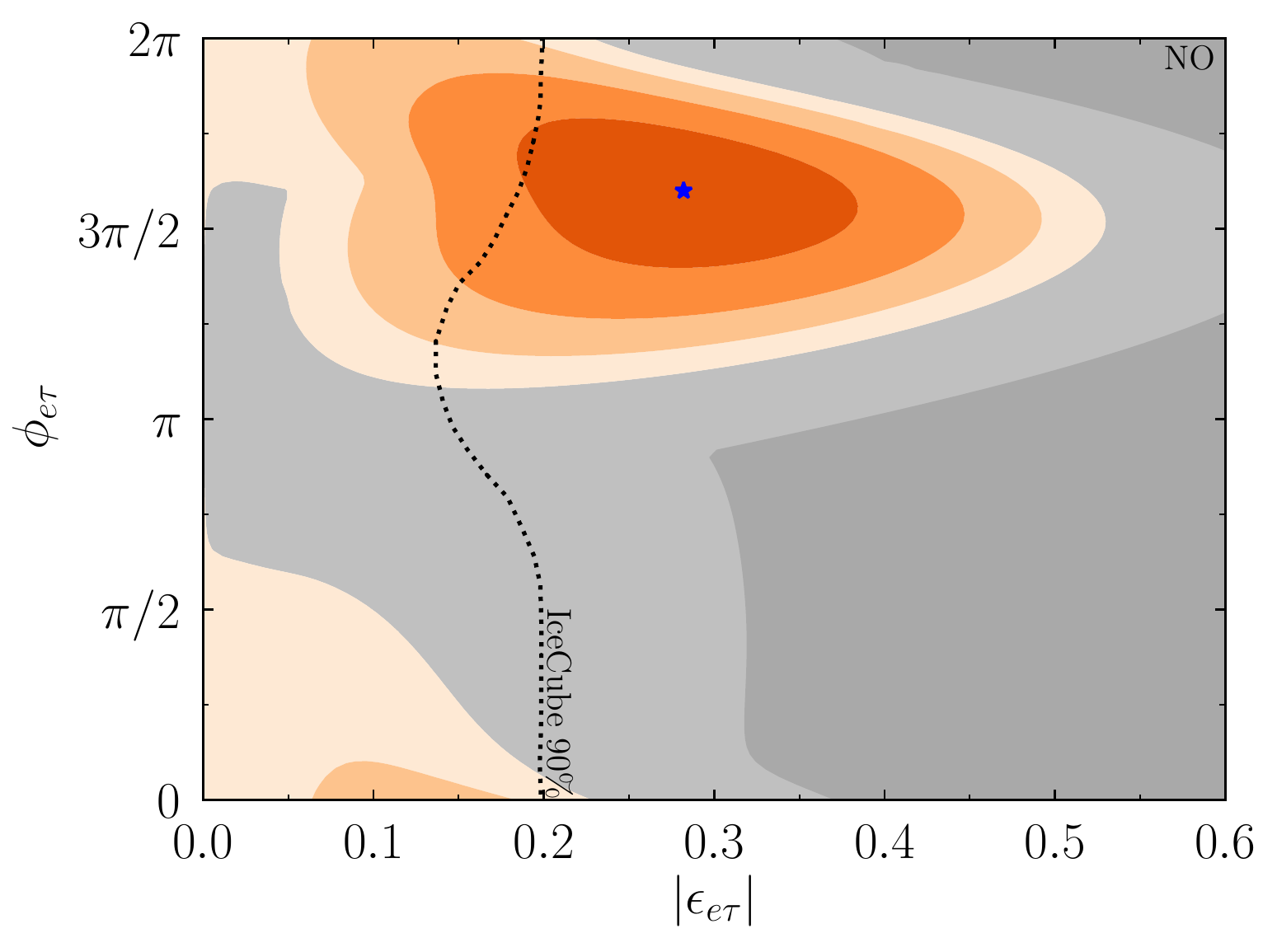}
\caption{The preferred parameter regions for $\epsilon_{e\tau}$ using the newest appearance and disappearance data from NOvA and T2K and assuming the NO.
The gray region is disfavored compared to the SM, and the dark gray region is ruled out by NOvA and T2K data at $\Delta\chi^2\leq-4.61$.
The blue star shows the best fit point.
Each of the orange contours are drawn at integer values of $\Delta\chi^2$.
See the table in the main text for the best parameters.
IceCube disfavors the region to the right of the black dotted curve at 90\% \cite{ICslides}.}
\label{fig:res_epset}
\end{figure}

We found that $\eps_{e\mu}$ explain the data well.
We also see in fig.~\ref{fig:res_epset} that $\eps_{e\tau}$ explains the data fairly well and is comparably allowable by other constraints.

\begin{figure*}
\centering
\includegraphics[width=0.49\textwidth]{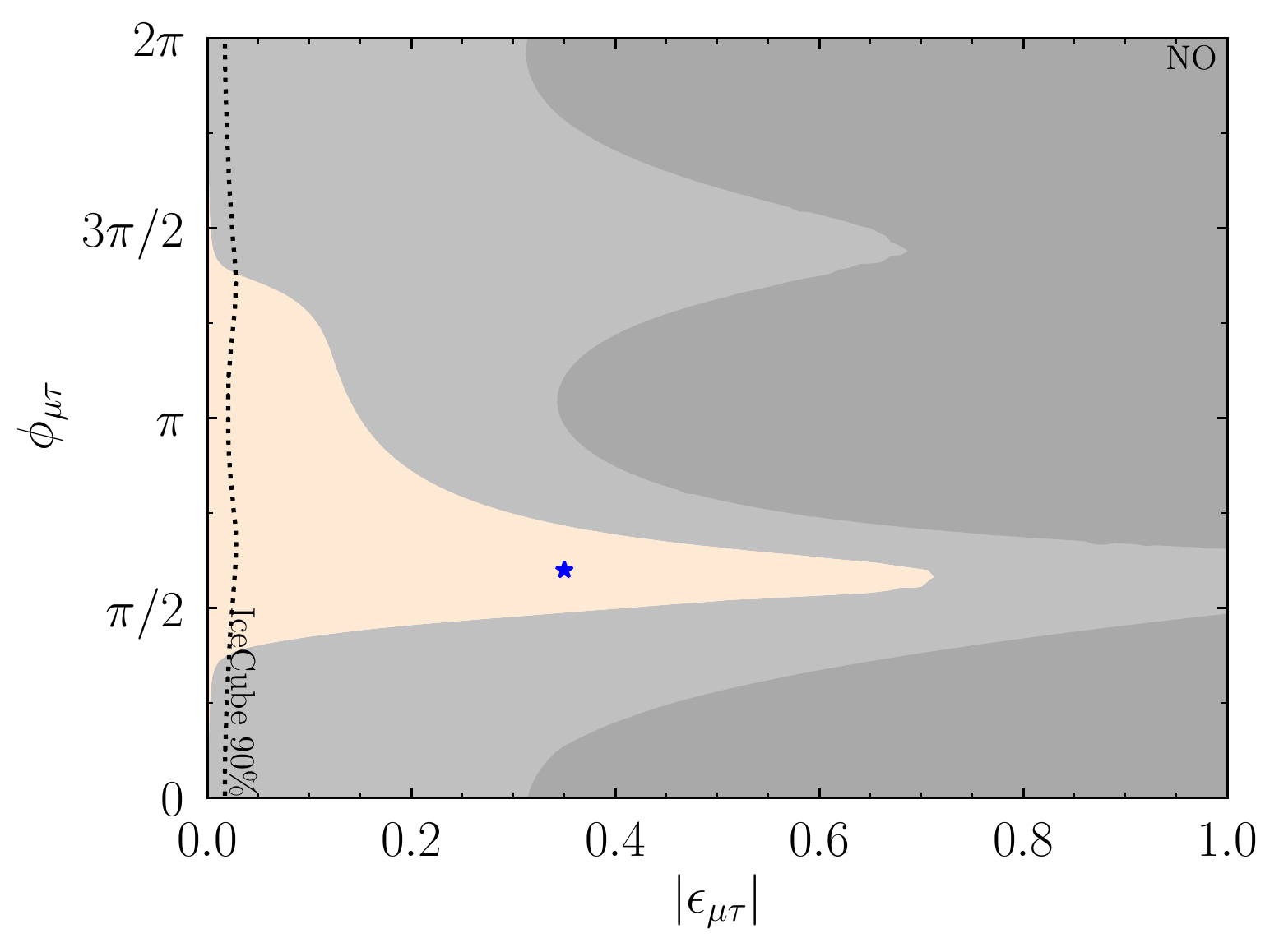}
\includegraphics[width=0.49\textwidth]{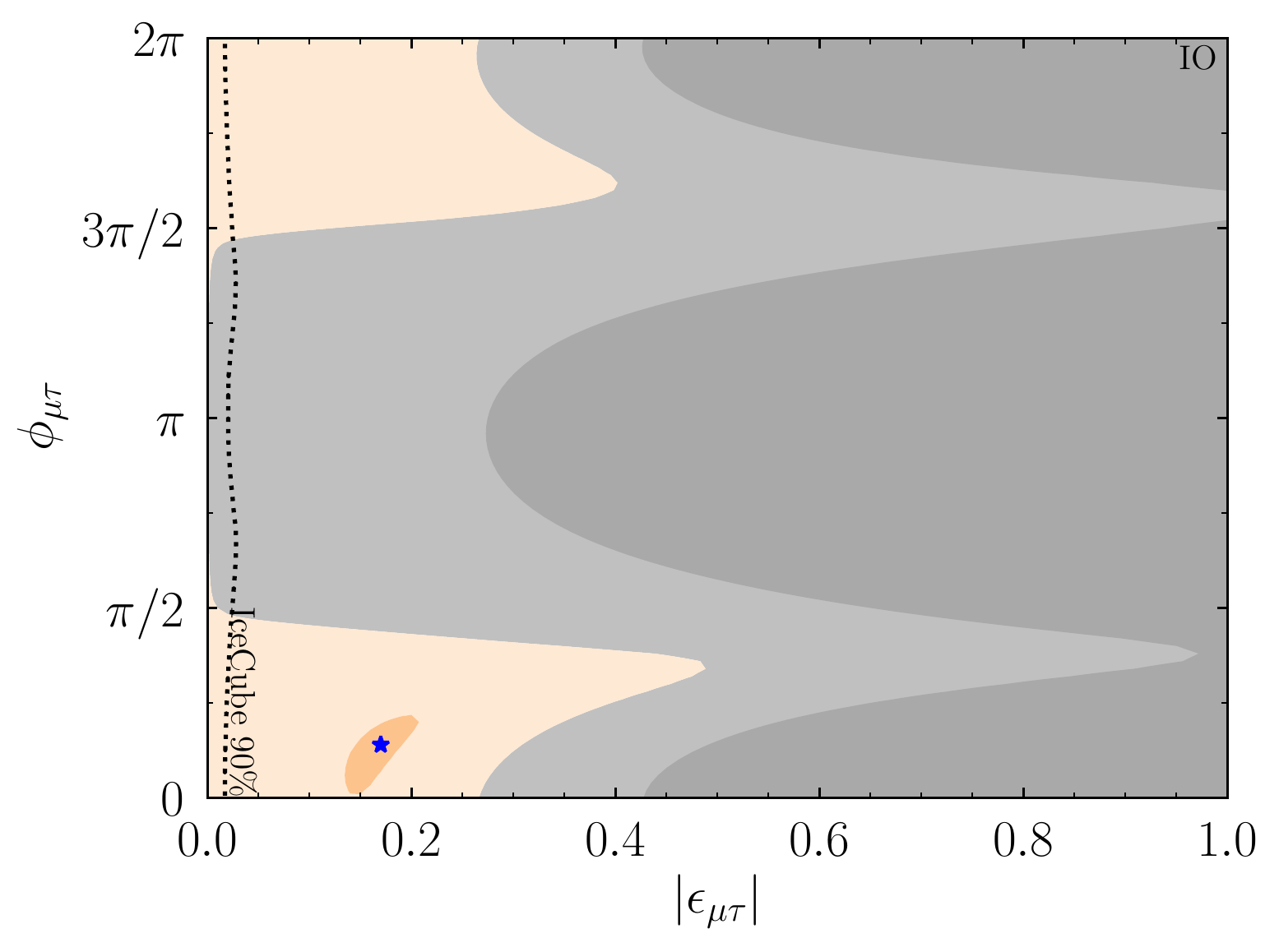}
\caption{The preferred parameter region for $\epsilon_{\mu\tau}$ using the newest appearance and disappearance data from NOvA and T2K and assuming the NO (left) or the IO (right).
The gray region is disfavored compared to the SM and the blue star shows the best fit point.
The orange contours are drawn at integer values of $\Delta\chi^2$.
See the table in the main text for the best parameters.
IceCube disfavors the region to the right of the black dotted curve at 90\% \cite{ICslides}.}
\label{fig:res_epsmt}
\end{figure*}

It is expected that $\eps_{\mu\tau}$ will not easily address the NOvA and T2K tension.
Moreover, there are very strong constraints on $\eps_{\mu\tau}$ from atmospheric data \cite{Esteban:2018ppq,Mitsuka:2011ty,ICslides}.
While these were generally derived under the assumption of real NSI, the relaxation to complex NSI should not significantly weaken the constraints.
Nonetheless, we show the preferred region in fig.~\ref{fig:res_epsmt} for NOvA and T2K data while marginalizing over $\theta_{23}$, $\Delta m^2_{31}$ (including a pull from Daya Bay), and $\delta$ while the other three standard oscillation parameters were set to their best fit values from Daya Bay and KamLAND.
SK has a bound on $\eps_{\mu\tau}$ that is slightly stronger than IceCube's at $\phi_{\mu\tau}=0,\pi$ \cite{Mitsuka:2011ty}, since this bound is only valid for CP conserving NSI, and since IceCube thoroughly rules out the regions of parameter space preferred by NOvA and T2K we do not show it on fig.~\ref{fig:res_epsmt}.

\end{document}